\documentclass[conference]{IEEEtran}
\IEEEoverridecommandlockouts
\usepackage{amsmath}
\usepackage{amsfonts}
\usepackage{amssymb}
\usepackage{mathtools}
\usepackage{amsthm}
\usepackage{fontenc}
\usepackage{graphicx}
\usepackage{algorithmic}
\usepackage{algorithm}
\usepackage[algo2e]{algorithm2e} 
\usepackage{booktabs}
\usepackage{subfigure}
\usepackage{tabulary}
\usepackage{booktabs}
\usepackage{siunitx}
\usepackage{caption}
\usepackage{comment}

\usepackage{cite}

\SetKw{KwBy}{by}
\usepackage[colorinlistoftodos]{todonotes}
\newcommand{\mv}[1]
{\mbox{\boldmath{$#1$}}}

\newcommand{\R}[1] {\textnormal{#1}}
 
\title{Learning for Semantic Knowledge Base-Guided Online Feature Transmission in Dynamic Channels}

\author{Xiangyu Gao, Yaping Sun, Dongyu Wei, Xiaodong Xu, Hao Chen,  Hao Yin, Shuguang Cui, \IEEEmembership{Fellow,~IEEE}
\thanks{X.~Gao, D.~Wei, H.~Yin are with the Department of Electrical and Computer Engineering, University of Washington, Seattle, USA. (email: \{xygao, wdy1223, haoyin\}@uw.edu). Y.~Sun and H.~Chen are with the Department of Broadband Communication, Peng Cheng Laboratory, Shenzhen 518000, China. (email: \{sunyp, chenh03\}@pcl.ac.cn). X.~Xu is with the Beijing University of Posts and Telecommunications, Beijing 100876, China, and affiliated with the Department of Broadband Communication, Peng Cheng Laboratory, Shenzhen 518000, China. (email: xuxiaodong@bupt.edu.cn). S.~Cui is with the School of Science and Engineering (SSE) and \textcolor{black}{the Future Network of Intelligent Institute (FNii)}, the Chinese University of Hong Kong (Shenzhen), Shenzhen 518172, China. S.~Cui is also with Shenzhen Research Institute of Big Data, Shenzhen 518172, China, and affiliated with the Department of Broadband Communication, Peng Cheng Laboratory, Shenzhen 518000, China (email: shuguangcui@cuhk.edu.cn).}
\thanks{Corresponding Author: Yaping Sun.}
}

\begin{document}
\maketitle

\begin{abstract}
With the proliferation of edge computing, efficient AI inference on edge devices has become essential for intelligent applications such as autonomous vehicles and VR/AR. In this context, we address the problem of efficient remote object recognition by optimizing feature transmission between mobile devices and edge servers. We propose an online optimization framework to address the challenge of dynamic channel conditions and device mobility in an end-to-end communication system. Our approach builds upon existing methods by leveraging a semantic knowledge base to drive multi-level feature transmission, accounting for temporal factors and dynamic elements throughout the transmission process. To solve the online optimization problem, we design a novel soft actor-critic-based deep reinforcement learning system with a carefully designed reward function for real-time decision-making, overcoming the optimization difficulty of the NP-hard problem and achieving the minimization of semantic loss while respecting latency constraints. Numerical results showcase the superiority of our approach compared to traditional greedy methods under various system setups.
\end{abstract}


\section{Introduction}

The resurgence of artificial intelligence (AI) has spurred the development of numerous intelligent applications, ranging from brain-to-computer interaction to autonomous driving \cite{ramp,gao2019experiments} and virtual/augmented reality (VR/AR) \cite{sun2019communications}. Ensuring AI inference capabilities on edge devices such as mobile phones, sensors, and vehicles is becoming increasingly crucial. In typical edge inference systems, data is transmitted from the edge devices to a physically proximate edge server (e.g., base station) for inference. This aims to alleviate the computational burden on the device and reduce service latency \cite{shao2022task}.

In this paper, our focus lies on a critical aspect of edge inference: efficient remote object recognition. Determining which information to transmit to the edge server is pivotal for minimizing transmission latency while maintaining satisfactory recognition performance. The conventional approach involves sending all input data. However, in typical mobile intelligent applications \cite{gao2021mimosar}, the sheer volume of collected data, including images and videos, leads to an impractical communication overhead \cite{shao2022task}. Prior study \cite{liu2019edge} introduced the dynamic region-of-interest encoding technique to compress raw frames for adaptive convolutional neural network (CNN) offloading in the mobile AR real-time object detection. Additionally, approaches presented in \cite{shao2021learning, guo2019distributed} involved extracting task-relevant features from raw input data and forwarding them to be processed by an edge server for task-oriented communication. However, these feature extractors rely on large-scale labeled training datasets and end-to-end deep neural network (DNN) training. Differing from the aforementioned literature, the work in \cite{sun2023zero} incorporated the knowledge base concept from semantic communication \cite{luo2022semantic}. It introduced multi-level feature transmission, driven by a semantic knowledge base (SKB), to facilitate remote zero-shot recognition and reduce dependence on extensive datasets. The term ``zero-shot'' emphasizes the ability to recognize novel image categories without any training samples. This is akin to how humans perform knowledge-based recognition, where they transfer their understanding to identify new classes solely based on textual descriptions \cite{sun2023semantic}. In this paper, we extend the concept of multi-level feature transmission for remote zero-shot recognition. However, we address it in an online optimization framework, factoring in the temporal dimension and accounting for dynamic channel quality. Due to the random motion of the mobile device, the channel between the device and server experiences unpredictable variations in transmission quality. As a result, the existing offline optimization algorithms \cite{sun2023zero, sun2023semantic} are not well-suited for our scenario.

Solving online optimization problems is challenging because we often lack complete or any knowledge of future events, necessitating us to make the best decision given the current circumstances. Reinforcement Learning (RL) has emerged as a powerful tool for addressing dynamic and uncertain environments, and it has found wide application in online scheduling and optimization tasks \cite{li2017deep, gao2023soft}. RL empowers an agent to learn from experience and adjust its actions over time to maximize rewards or achieve the best possible outcome. For instance, in \cite{gao2023joint}, a Soft Actor-Critic (SAC) approach was introduced to jointly optimize computing, pushing, and caching policies in mobile edge computing (MEC) networks for mobile users in motion. This resulted in reductions in both transmission and computation costs. Similarly, in \cite{yin2022routing}, an RL framework was proposed to tackle resource allocation and routing challenges in integrated access and backhaul (IAB) multi-hop networks.

This paper tackles the challenge of online optimization in remote object detection by introducing an innovative SKB-powered multi-level feature transmission system. Our contributions are outlined below:
\begin{itemize}
    \item First, we present a model for a mobile device-server communication system designed for remote object detection, incorporating dynamic elements such as moving mobile devices and fluctuating transmission channels. This is formulated as an online optimization problem with the dual objectives of minimizing semantic loss and adhering to specified transmission latency thresholds.
    \item Then, to address the online optimization challenge, which inherently resembles a multi-choice knapsack problem known to be NP-hard \cite{multichoice}, we propose a SAC-based DRL system. This system provides real-time multi-level feature transmission decisions at each time slot. We have designed a specific reward structure to guide the decision-making process while ensuring latency constraints are met.
    \item Finally, the presented numerical results demonstrate the remarkable performance improvements achieved by our proposed design across various system configurations, in comparison to two baseline greedy approaches.
\end{itemize}

\begin{figure}[t]
    \centering   
    \includegraphics[width=.45\textwidth]{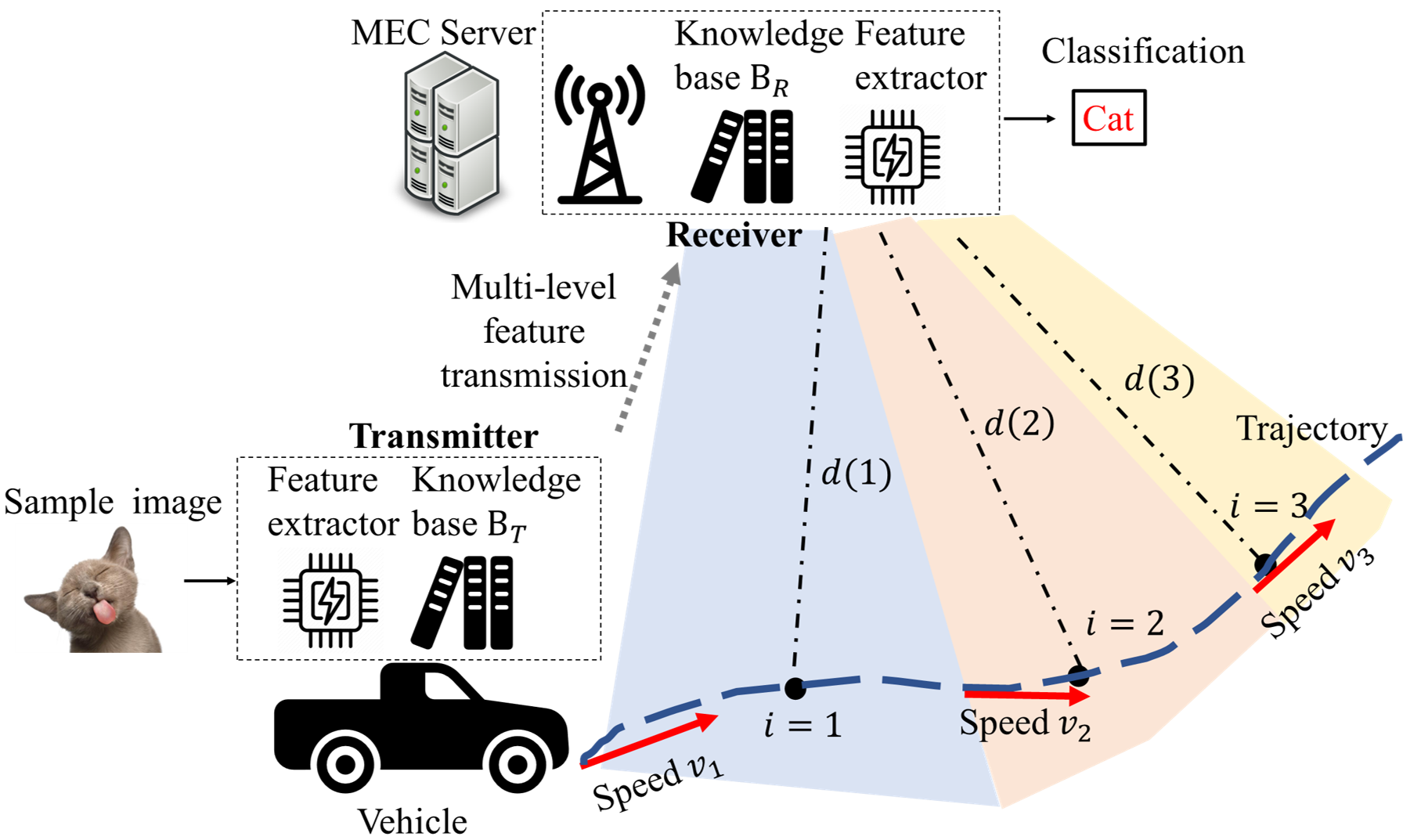}
   \caption{A SKB-enabled E2E communication system for remote object recognition.} \label{fig:e2e}
\end{figure}

\section{System Model}
As depicted in Fig.~\ref{fig:e2e}, we consider a novel end-to-end (E2E) communication system between a vehicle and an MEC server. Both the vehicle and MEC server function as the transmitter and receiver, equipped with a unique SKB and a multi-level feature extractor. At time $i$ ($i=1,2,3,...$), the vehicle extracts features from sample images and transmits them to the MEC server for remote object classification. The vehicle maintains a constant speed $v$ for a short duration (e.g., 50 frames) and then undergoes random speed changes. The distance between the vehicle and the MEC server at time $i$ is denoted as $d(i)$, and its long-term predictability is affected by the random variations in vehicle speed. This distance plays a crucial role in determining the quality of the communication link.

\subsection{Semantic knowledge base (SKB)}
The SKB of each mobile device is characterized by a collection of semantic vectors, each representing distinct classes of information stored on the device. \cite{sun2023zero}. Leveraging the local SKB, a mobile device can identify attributes associated with these classes and discern the semantic relationships among them. To formalize this, let $\mathcal{C} \triangleq \left\{1,2,\cdots,C\right\}$ denote the set of all classes, and $\mathcal{S} \triangleq \left\{\mv s_1, \mv s_2, \cdots, \mv s_C\right\}$ represent the set of semantic vectors corresponding to these classes. These vectors are obtained from Word2Vec or can be manually annotated semantic attribute vectors \cite{kodirov2017semantic}. Next, we define the SKB for both the transmitter and the receiver:
\begin{itemize}
    \item SKB at the transmitter: Let $t_c \in \{0,1\}$ indicate the semantic knowledge indicator of class $c$ at the transmitter. Here, $t_c = 1$ signifies that the transmitter possesses the semantic information for class $c$, meaning the semantic vector $\mv s_c$ is stored at the transmitter; conversely, $t_c = 0$ indicates otherwise. Denote by $\mathcal{B}_T \triangleq \{c\in \mathcal{C}: t_c = 1\}$ the set of class labels for which the semantic vectors are stored at the transmitter, constituting the SKB at the transmitter \cite{sun2023zero}.
    \item SKB at the receiver: Similarly, let $r_c \in \{0,1\}$ denote the semantic knowledge indicator of class $c$ at the receiver. Define $\mathcal{B}_R \triangleq \{c\in \mathcal{C}: r_c = 1\}$ as the set of class labels for which the semantic vectors are stored at the receiver, forming the SKB at the receiver. 
\end{itemize}

\subsection{Multi-level Feature Extractor} 

Consider a dataset $\mathcal{N}$ comprising $N$ training image samples, each represented by its features $\left(\mv V, \mv C, \mv S\right)$. Specifically: $\mv V \in \mathbb{R}^{D_v \times N}$ represents the $D_v$-dimensional visual feature matrix of the $N$ image samples. These features are extracted using pre-trained deep convolutional neural networks (CNNs) \cite{ramp}. For instance, they can be obtained before the fully connected layer. $\mv C \triangleq (c_n)_{n\in \mathcal{N}} \in \mathcal{C}_{\text{Tr}}^{N}$ denotes the class label vector of the image samples. Here, $c_n \in \mathcal{C}_{\text{Tr}}$ refers to the class label of sample $n$, and $c_n \in \mathcal{C}_{\text{Tr}}$ encompasses the set of classes observed in the training samples. $\mv S \in \mathbb{R}^{D_s \times N}$ represents the $D_s$-dimensional semantic feature matrix. Each column corresponds to the semantic feature vector of class $c_n$ for sample $n$. Using the training samples $\mathcal{N}$ and their associated features $\left(\mv V, \mv C, \mv S\right)$, we can obtain the visual encoder $\mv P_{v}$, visual decoder $\mv P_{v}^T$, semantic encoder $\mv P_{s}$, and semantic decoder $\mv P_{s}^T$ according to the multi-level feature extractor proposed in \cite{sun2023zero}. These components facilitate the extraction of features for unseen testing samples.

With $\mv P_{v}$, $\mv P_{v}^T$, $\mv P_{s}$, and $\mv P_{s}^T$, we construct below multi-level feature extractor from \cite{sun2023zero} for any given testing sample $\mv l$:
\begin{itemize}
    \item 1st-level visual feature, denoted as $\mv v = f(\mv l) \in \mathbb{R}^{D_v}$, is derived by applying the pre-trained CNN model $f(\cdot)$ to the image sample $\mv l$.
    \item 2nd-level intermediate feature, denoted as $\mv f = \mv P_v \mv v \in \mathbb{R}^k$, is derived by using the visual encoder $\mv P_v$ to project the visual feature $\mv v$.
    \item 3rd-level semantic feature, denoted as $\mv s = \mv P_s^T \mv f \in \mathbb{R}^{D_s}$, is derived by applying the semantic decoder $\mv P_s^T$ to the intermediate feature $\mv f$.
    \item 4th-level estimated class label, expressed as $\hat{c} = \arg \min_{c\in \mathcal{C}} \Arrowvert \mv s_c - \mv s \Arrowvert^2_F$, where $\mathcal{C}$ represents the available class set, and $\mv s_c$ refers to the semantic vector for class $c$.
\end{itemize}

\subsection{Multi-level Feature Transmission Policy}
Both the transmitter and receiver are equipped with their own multi-level feature extractors, which are trained from their respective datasets denoted as $(\mv V_t, \mv C_t, \mv S_t)$ and $(\mv V_r, \mv C_r, \mv S_r)$. Specifically: For the transmitter, we use $\mv P_{t,v}$, $\mv P_{t,v}^T$, $\mv P_{t,s}$, and $\mv P_{t,s}^T$ to represent the visual encoder, visual decoder, semantic encoder, and semantic decoder, respectively. Similarly, for the receiver, we use $\mv P_{r,v}$, $\mv P_{r,v}^T$, $\mv P_{r,s}$, and $\mv P_{r,s}^T$ with the same meanings as for the transmitter. 

At any given time $i$ with an inter-distance between the transmitter and receiver denoted as $d(i)$, there exist $M$ testing samples, denoted as $\mathcal{M} \triangleq \{1,2,\cdots, M\}$, which have not been encountered during the training process at both the transmitter and receiver. Utilizing the previously described multi-level feature extractor, we have the following four types of transmission choices \cite{sun2023zero} at the transmitter for each testing sample in order to complete the classification task. Let $x_{m,l}(i) \in \{0,1\}$ represent the feature transmission choice of a sample at time $i$, where $x_{m,l}(i)=1$ indicates that the $l$-th level feature vector of sample $m$ is transmitted. To ensure the delivery of semantic information for each sample to the receiver, we enforce the constraint: $\sum_{l=1}^4 x_{m,l}(i)= 1$, for all $m \in \mathcal{M}$ and all $i$. 

\begin{itemize}
    \item  1st-level visual feature transmission: When $x_{m,1}(i) =1$, the visual feature vector of sample $m$, represented by $\mv v_m \in \mathbb{R}^{D_v}$, is transmitted to the receiver. Based on \cite{sun2023zero}, the receiver estimates the class using its multi-level feature extractor and SKB, i.e., $\hat{c}_{r,m} = \arg \min_{c\in \mathcal{B}_R} \Arrowvert \mv s_c - \mv P_{r,s}^T \mv P_{r,v}\mv v_m \Arrowvert^2_F$. The corresponding semantic loss is given by $L_{m,1}(i) = \min_{c\in \mathcal{B}_R} \Arrowvert \mv s_{c} - \mv P_{r,s}^T \mv P_{r,v}\mv v_m \Arrowvert^2_F$. The required transmission latency is $T_{m,1}(i) = \frac{D_vQ}{R(i)}$, where $Q$ denotes the quantization level \cite{shao2021learning}, and $R(i)$ is the achievable data rate from the transmitter to the receiver at time $i$. The achievable data rate $R(i)$ is determined by the distance $d(i)$ between the transmitter and receiver, overall available power $P_U$, bandwidth $B$, and noise power spectral density $N_0$. The term $\beta_0 \left(\frac{d(i)}{d_0}\right)^{-\zeta}$ represents the path loss, where $\beta_0$ is the path loss at the reference distance $d_0$, and $\zeta$ is the path loss exponent.
    \begin{equation}
        R(i) = B \log_2{\left( 1 +  P_U\beta_0 \left(\frac{d(i)}{d_0}\right)^{-\zeta}\frac{1}{BN_0}\right)}
    \end{equation}
    
    \item 2nd-level intermediate feature transmission: Consider $x_{m,2}(i)=1$, the intermediate feature vector of sample $m$ at time $i$, represented by $\mv f_m = \mv P_{t,v}\mv v_m \in \mathbb{R}^{k}$, is transmitted to the receiver. The receiver estimates the class based on its semantic decoder and SKB \cite{sun2023zero}, i.e., $\hat{c}_{r,m} = \arg \min_{c \in \mathcal{B}_R} \Arrowvert \mv s_c - \mv P_{r,s}^T \mv f_m \Arrowvert^2_F$. The corresponding semantic loss is $ L_{m,2}(i) = \min_{c\in \mathcal{B}_R} \Arrowvert \mv s_{c} - \mv P_{r,s}^T \mv P_{t,v}\mv v_m \Arrowvert^2_F$. The required transmission latency is $T_{m,2}(i) = \frac{kQ}{R(i)}$. 
    
    \item 3rd-level semantic feature transmission: When $x_{m,3}(i)=1$, we transmit the semantic feature vector of sample $m$, represented by $\mv s_m = \mv P_{t,s}^T\mv P_{t,v}\mv v_m \in \mathbb{R}^{D_s}$, to the receiver. The receiver estimates the class based on its SKB \cite{sun2023zero}, i.e., $\hat{c}_{r,m} = \arg \min_{c\in \mathcal{B}_R} \Arrowvert \mv s_c - \mv s_m \Arrowvert^2_F$. The corresponding semantic loss is $ L_{m,3}(i) = \min_{c\in \mathcal{B}_R} \Arrowvert \mv s_{c} - \mv P_{t,s}^T \mv P_{t,v}\mv v_m \Arrowvert^2_F$. The required transmission latency is $T_{m,3}(i) = \frac{D_sQ}{R(i)}$.
   
    \item 4th-level estimated class knowledge transmission: When $x_{m,4}(i)=1$, the transmitter will estimate the class of the image sample, represented as $\hat{c}_{t,m} = \arg \min_{c\in \mathcal{B}_T} \Arrowvert \mv s_c - \mv P_{t,s}^T\mv P_{t,v} \mv v_m \Arrowvert^2_F$. The estimated label $\hat{c}_{t,m}$ is then transmitted to the receiver. The transmission load of the estimated label is $Q$. Thus, the required transmission latency is $T_{m,4}(i) = \frac{Q}{R(i)}$. The corresponding semantic loss is $L_{m,4}(i) = \min_{c\in \mathcal{B}_T} \Arrowvert  \mv s_{c} - \mv P_{t,s}^T \mv P_{t,v}\mv v_m \Arrowvert^2_F$.
\end{itemize}

To summarize, the average semantic loss deemed by the multi-level feature transmission for $M$ samples at time $i$ is given by
\begin{align}
    \frac{1}{M}\sum_{m=1}^M \sum_{l=1}^4 x_{m,l}(i) L_{m,l}(i),
\end{align}
and the average transmission latency is given by
\begin{align}
   \frac{1}{M} \sum_{m=1}^M \sum_{l=1}^4 x_{m,l}(i) T_{m,l}(i).
\end{align}

\section{Problem Formulation}

Given this setup, our objective is to minimize the semantic loss over time, leveraging the knowledge of both the transmitter and receiver, through optimization of the multi-level feature transmission policy $\mv x \triangleq \left(x_{m,l}\right)_{m\in \mathcal{M},l\in {1,2,3,4}}$. The semantic loss optimization is subject to a constraint on the average transmission latency, which is bounded by a threshold $\tau$. The optimization problem can be formulated as:
\begin{align}
    (\text{P1})\ & \min_{\mv x}\   \limsup_{T\rightarrow \infty} \frac{1}{T}\frac{1}{M}\sum_{i=1}^{T} \sum_{m=1}^M \sum_{l=1}^4 x_{m,l}(i) L_{m,l}(i)  \\
    &\ s.t. \ \limsup_{T\rightarrow \infty} \frac{1}{T}\frac{1}{M} \sum_{m=1}^M \sum_{l=1}^4 x_{m,l}(i)T_{m,l}(i) \leq \tau \label{eq:taucons}\\
    &\ \ \ \ \  \ \sum_{l=1}^4 x_{m,l}(i) = 1, \forall m \in \mathcal{M}, i\in \{1,\dots, T\} \label{eq:sum1}\\
    &\ \ \ \ \ \ \  x_{m,l} (i) \in \{0,1\}, \forall m \in \mathcal{M}, l \in \{1,2,3,4\}\label{eq:binary}
\end{align}

Problem~(P1) can be observed as an online multi-choice knapsack problem, known to be NP-hard \cite{multichoice}. To ensure problem~(P1) is feasible, it is assumed that $\frac{Q}{R(i)} \leq \tau$ holds throughout this paper. Let $\mv x^* \triangleq (x^*_{m,l})_{m\in \mathcal{M},l \in \{1,2,3,4\}}$ denote the optimal multi-level transmission policy for problem~(P1). To find $\mv x^*$, we employ a soft actor-critic-based DRL algorithm.

\section{Soft Actor-Critic (SAC)-Based Solver}
\subsection{System State and Action for SAC} \label{sec:SAC_action}

At time $i$, the system state $\mv{a}(i)$ for SAC is designed as $\mv{a}=\left(\left(L_{m,l}(i), T_{m,l}(i)\right)_{m \in \mathcal{M}, l \in \{1,2,3,4\}}, L_{\R{avg}}(i), T_{\R{avg}}(i), v(i), i\right)$, with a vector size of $4M+4$. The values of $L_{\text{avg}}(i)$ and $T_{\text{avg}}(i)$ are calculated by averaging all the historical semantic losses and transmission latencies up to time $i$.

While the SAC algorithm is designed to solve continuous-action problems, the multi-level transmission policy $\mv{x}$ in our formulated problem is discrete. To align these, we define the system action of the SAC as $\mathbf{x}=\left(\Bar{x}_m \in \left[0.5, 4.5\right], m=1, 2, \dots, M \right)$. Here, $\Bar{x}_m$ values within the ranges $[0.5, 1.5)$, $[1.5, 2.5)$, $[2.5, 3.5)$, and $[3.5, 4.5]$ are mapped to $x_{m, 1}=1$, $x_{m, 2}=1$, $x_{m, 3}=1$, and $x_{m, 4}=1$ respectively. After obtaining the continuous action $\mathbf{x}$ from SAC, we apply this mapping to obtain the discrete action $\mv{x}$, which is then used to update the system state and reward.

\subsection{SAC Learning}
\label{sec:sac_learning}
SAC is an off-policy deep reinforcement learning method that retains the benefits of entropy maximization and stability while offering sample-efficient learning \cite{haarnoja2018soft}. It operates within an actor-critic framework where the actor's role is to maximize the expected reward while also maximizing entropy. The critic, on the other hand, assesses the effectiveness of the policy being followed. The objective of maximum-entropy Reinforcement Learning (RL) is defined as:
\begin{equation}
    J(\pi)=\sum_{i=0}^T \mathbb{E}_{\left(\mathbf{a}(i), \mathbf{x}(i)\right)}\left[r\left(\mathbf{a}(i), \mathbf{x}(i)\right)+\alpha \mathcal{H}\left(\pi\left(\mathbf{x}(i)\mid \cdot \right)\right)\right]
    \label{eq:ge_rl}
    \nonumber 
\end{equation}
where $\mathcal{H}\left(\pi\left(\mathbf{x}(i)\mid \cdot \right)\right)=\mathbb{E}_{\mathbf{a}(i)}\left[-\log \pi\left(\mathbf{x}(i) \mid \mathbf{a}(i) \right)\right]$ represents the entropy term. Here, $\pi(\cdot)$ denotes the policy. $\alpha$ is the temperature parameter that determines the relative importance of the entropy term compared to the reward $r(\mathbf{a}(i), \mathbf{x}(i))$ for state $\mathbf{a}(i)$ and action $\mathbf{x}(i)$ at time $i$.

The SAC algorithm employs a policy iteration approach to address maximum-entropy Reinforcement Learning (RL). This approach involves two key components: the soft Q-function $Q_{\theta}\left( \mathbf{a}(i), \mathbf{x}(i)\right)$, and the policy $\pi_{\phi}\left(\mathbf{a}(i) \mid \mathbf{x}(i)\right)$. Here, $\theta$ and $\phi$ represent parameters used in approximating neural networks, which are crucial for handling large continuous domains \cite{gao2023soft, gao2023joint}. Additionally, SAC incorporates a target soft Q-function $Q_{\bar{\theta}}$ with parameters $\bar{\theta}$, which is derived as an exponentially moving average of the parameter $\theta$. This inclusion aids in stabilizing the training process. The update rules for $\theta$ and $\phi$, as well as the complete SAC learning algorithm, are detailed in Algorithm~\ref{alg:alg1}, following the guidelines presented in \cite{haarnoja2018soft}. The step sizes (or learning rates) for stochastic gradient descent, denoted as $\lambda_Q$ and $\lambda_\pi$, are set to $1\times10^{-4}$. The target smoothing coefficient $\xi$ is chosen to be 0.005. Moreover, $\mathcal{D}$ represents the distribution of previously sampled states and actions, $p$ signifies the transition probability between states, and $\Delta(\cdot)$ denotes the gradient function.

\subsection{Reward Design}
The reward $r(\mathbf{a}(i), \mathbf{x}(i))$ in the SAC framework, given state $\mathbf{a}(i)$ and action $\mathbf{x}(i)$, is a composite of the semantic loss and a penalty for transmission latency. Assuming action $\mathbf{x}(i)$ corresponds to $x_{m, \hat{l}}(i)=1$, the reward is calculated as:
\begin{align}
\nonumber
r(\mathbf{x}(i), \mathbf{a}(i)) = - \frac{1}{M} \sum_{m=1}^M L_{m,\hat{l}}(i) -  \kappa_1 Y_{1,m}(i) -  \kappa_2 Y_{2,m}(i).
\end{align}
Here, $Y_{1,m}(i)$ and $Y_{2,m}(i)$ are penalty terms:
\begin{align}
& Y_{1,m}(i) = \delta({T_{\R{avg}} > \tau}) \cdot (T_{m,\hat{l}}(i) - \min_{l \in \{1,2,3,4\}}T_{m,l}(i))\\
& Y_{2,m}(i) = \delta({T_{\R{avg}} \le \tau}) \cdot (\tau - T_{\R{avg}}(i))
\end{align}

\noindent where $Y_{1,m}$ represents the penalty in cases where the average transmission latency exceeds the threshold $\tau$, and $Y_{2,m}$ in cases where it is below the threshold. The $\delta(\cdot)$ function ensures that only the appropriate term is activated. Additionally, $\kappa_1$ and $\kappa_2$ are weighting coefficients for $Y_{1,m}$ and $Y_{2,m}$ respectively. For this paper, we have chosen $\kappa_1=500$ and $\kappa_2=500$.

\begin{algorithm}
\caption{SAC Learning \cite{gao2023soft}}
\label{alg:alg1}
Initialize parameters $\theta, \bar\theta, \phi$ for networks $Q_{\theta}$, $Q_{\bar\theta}$, $\pi_{\phi}$\;

Initialize learning rate $\lambda_Q$, $\lambda_\pi$, and weight $\xi$\;

\For{each episode}{
    \For{each environment step}{
    $\mathbf{a}(i) \sim \pi_\phi\left(\mathbf{x}(i) \mid \mathbf{a}(i) \right)$\;
    
    $\mathbf{x}(i+1) \sim p\left(\mathbf{x}(i+1) \mid \mathbf{a}(i), \mathbf{x}(i)\right)$\;
    
    calculate $r\left(\mathbf{a}(i), \mathbf{x}(i)\right)$\;
    
    $\mathcal{D} \leftarrow \mathcal{D} \cup\left\{\left(\mathbf{a}(i), \mathbf{x}(i), r\left(\mathbf{a}(i), \mathbf{x}(i)\right), \mathbf{x}(i+1)\right)\right\}$\;
    }
    
    \For{each gradient step}{
    $\theta_i \leftarrow \theta_i-\lambda_Q \hat{\nabla}_{\theta_i}J_Q\left(\theta_i\right) \text { for } i \in\{1,2\}$\;
    
    $\phi \leftarrow \phi-\lambda_\pi \hat{\nabla}_\phi J_\pi(\phi)$\;
    
    $\bar \theta_i \leftarrow \xi \theta_i + (1-\xi) \bar \theta_i \text { for } i \in\{1,2\}$\;
    }
}
\end{algorithm}

\section{Simulation and Analysis}
\subsection{Baselines}
The proposed SAC solver is called ``DRL'' in the following analysis. We implemented two online baselines for comparison with DRL: the Loss-First Greedy Algorithm and the Latency-First Greedy Algorithm.

\subsubsection{Loss-First Greedy Algorithm}
At time $i$, we select the instance $m$ with the lowest semantic loss. If the average history transmission latency $L_{\R{avg}} (i)$ does not exceed the constraint $\tau$, we include instance $m$ in the output sequence. Otherwise, we opt for the 4th-level estimated class label transmission.

\subsubsection{Latency-First Greedy Algorithm}
At time $i$, we prioritize choosing the instance $m$ with the lowest latency.

\subsection{Implementation}
The Animals with Attributes (AwA) dataset \cite{awa} is adopted. First, there are $24295$ samples from the AwA dataset used for training the semantic knowledge base and getting the multi-level feature extractors. Second, there are $6180$ unseen samples of data used for DRL training and testing. Specifically, $80\%$ of them are used for DRL training, and $20\%$ are used for DRL testing and baseline evaluation. The $20\%$ data will not be memorized or cached by the DRL system. 

In our simulation, we considered the number of classes in the transmitter knowledge space to be $5$ and the number of classes in the receiver knowledge space to be $10$. The path loss gain is modeled by $\beta_0 = \SI{-30}{dB}$, $d_0 = \SI{1}{m}$, $\zeta = 2$. The transmission rate $R$ is set by $B = \SI{100}{KHz}$, $N_0 = \SI{-114}{dBm/Hz}$, and $P_U = \SI{10}{dBm}$. The vehicle's velocity is changed randomly every 50 timeslots (i.e., when $(i-1)\% 50 =0$). The velocity range is $[\SI{-6}{m/s}, \SI{6}{ms}]$. The default transmission latency constraint is \SI{3}{ms}. The number of samples for the timeslot is $M=1$.

The implementation of the system is done with Python and PyTorch. The training and testing were deployed on a PC with TITAN RTX GPU using batch size 256, discount factor $\gamma=0.99$, automatic entropy temperature $\alpha$ tuning \cite{haarnoja2018soft}, network hidden-layer size 256, one model update per step, one target update per 1000 steps, and replay buffer size of $1\times10^7$, learning rate $1\times10^{-4}$. We make 1 testing epoch after every 5 training epochs and stop the training and testing when the reward and loss converge. 

\subsection{DRL Convergence}
\begin{figure}
\centering
\includegraphics[width=0.48\textwidth]{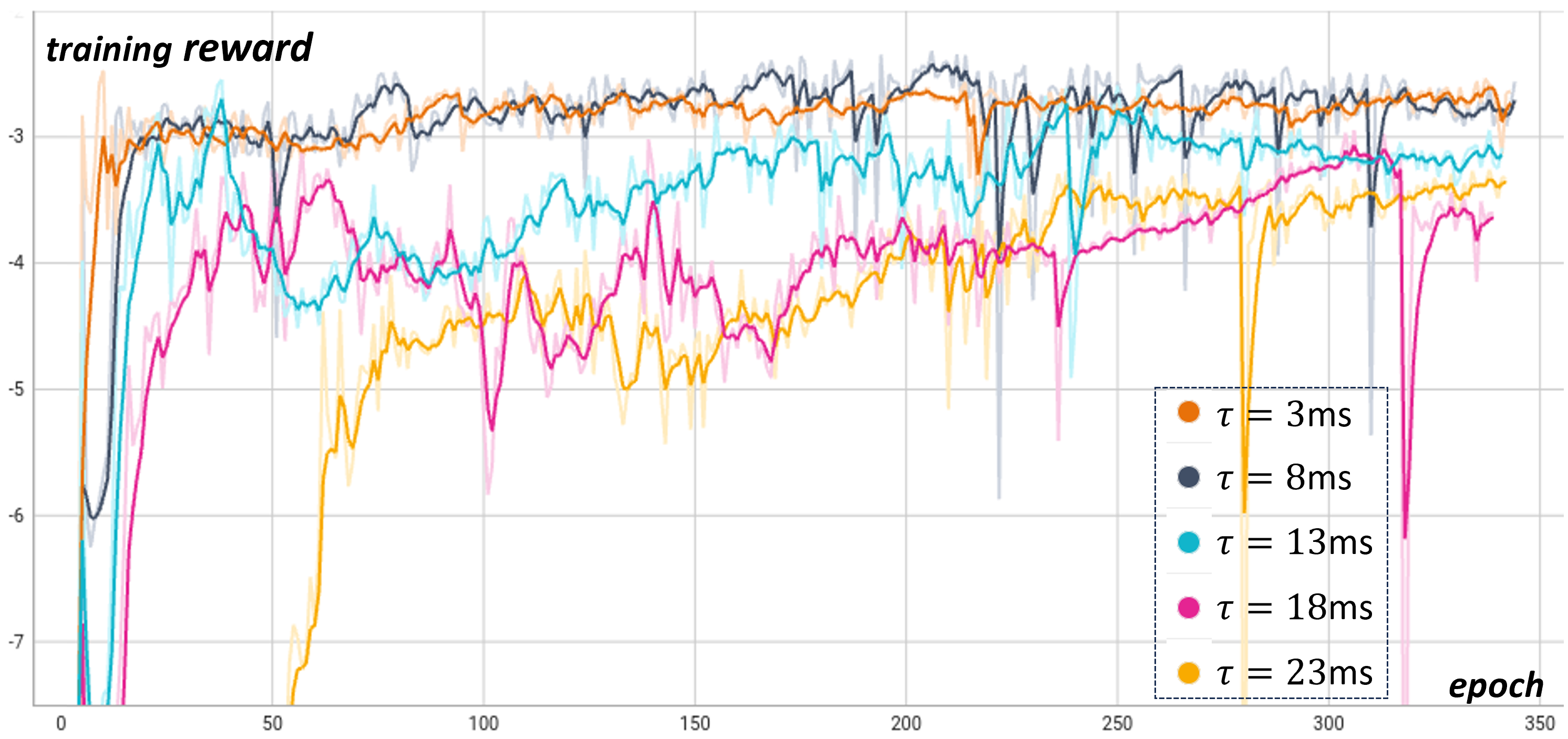}
  \caption{Training convergence of DRL algorithm for different configuration cases of transmission latency constraint $\tau$.}
  \label{fig:converge}
\end{figure}

The convergence of the DRL algorithm is illustrated in Fig.~\ref{fig:converge}, showcasing training rewards plotted against epochs for various $\tau$ constraint configurations. Notably, for smaller $\tau$ values like \SI{3}{ms} and \SI{8}{ms}, the DRL algorithm converges swiftly in approximately 100 epochs. Conversely, more intricate configurations with larger $\tau$ values, such as \SI{13}{ms}, \SI{18}{ms}, and \SI{23}{ms}, generally necessitate 300 epochs or more to reach convergence.

\subsection{Metrics}
The evaluation of the proposed algorithm and baselines involves three key metrics: average semantic loss, average transmission latency, and accuracy.

Average semantic loss and average transmission latency are determined by computing the mean of the semantic loss and transmission latency values across all testing samples and time points. In both cases, lower values indicate better performance. Accuracy is assessed by examining the classification outcomes at the receiver end. The accuracy calculation involves several steps: for 1st, 2nd, and 3rd-level feature transmissions, we compare the receiver's estimated class $\hat{c}_{r,m}$ with the ground truth. In the case of 4th-level feature transmission, we compare the transmitter's estimated class $\hat{c}_{t,m}$ with the ground truth. Accuracy represents the proportion of correctly predicted data points relative to the total number of data points. Higher accuracy values reflect superior performance.

\subsection{Numerical Results}

\subsubsection{Varying the transmission latency constraint $\tau$}

The average semantic loss, average transmission latency, and accuracy of the proposed DRL algorithm, along with the Loss-First Greedy and Latency-First Greedy algorithms, were compared under different latency constraints $\tau =$ \SI{3}{ms}, \SI{8}{ms}, \SI{13}{ms}, \SI{18}{ms}, \SI{23}{ms}, and \SI{28}{ms}, as shown in Fig.~\ref{fig:vary_tau}(a)-(c). The DRL algorithm consistently outperforms both baselines, achieving significantly smaller average semantic loss and higher accuracy across all setups. As $\tau$ increases, the average semantic loss for DRL continues to decrease (see Fig.~\ref{fig:vary_tau}(a)). This is because lower-level features, which require more transmission latency, can now be transmitted. The increase in accuracy is attributed to the fact that transmitting low-level information incurs less semantic loss during training (refer to Fig.~\ref{fig:vary_tau}(c)). Furthermore, the average transmission latency for DRL consistently remains below the constraint threshold, as depicted in Fig.~\ref{fig:vary_tau}(b). In contrast, the Loss-First Greedy algorithm consistently exhibits slightly higher average transmission latency than the specified constraint $\tau$. The evaluation results for Latency-First Greedy remain consistent, as the constraint has no effect on action selection, and it consistently opts for the lowest-level feature transmission. Although the Loss-First Greedy algorithm shows lower semantic loss as $\tau$ increases, this improvement is not on par with the performance achieved by the DRL algorithm. This discrepancy arises from the fact that the greedy algorithm lacks the intelligence to adjust actions in response to complex system states and rapidly changing channel conditions.

\subsubsection{Varying transmitter SKB size $B_K$}
\begin{figure}
\centering
\subfigure[]{\includegraphics[width=0.24\textwidth]{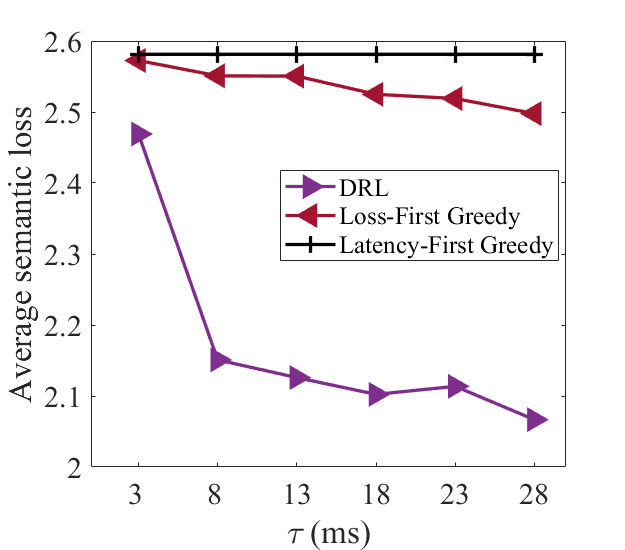}}
\subfigure[]{\includegraphics[width=0.24\textwidth]{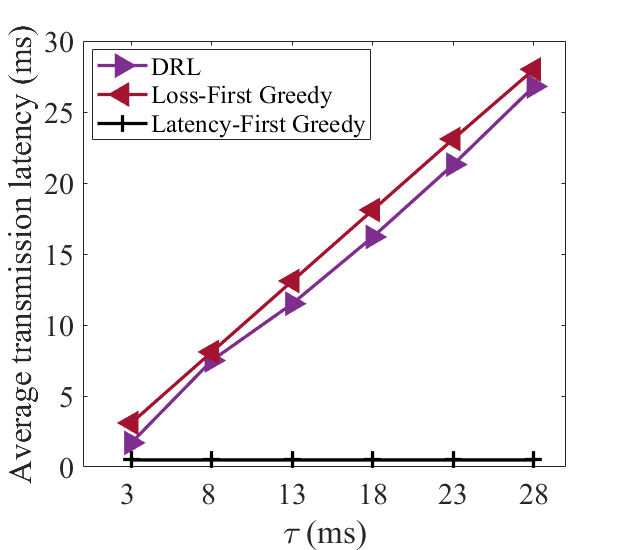}}
\subfigure[]{\includegraphics[width=0.24\textwidth]{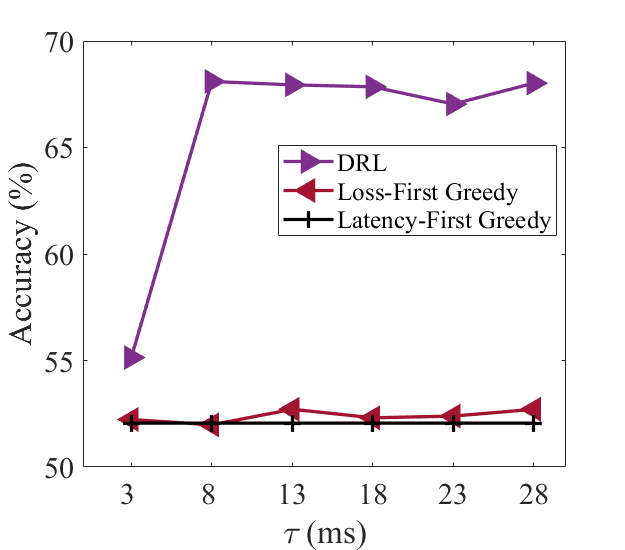}}
\subfigure[]
{\includegraphics[width=0.24\textwidth]{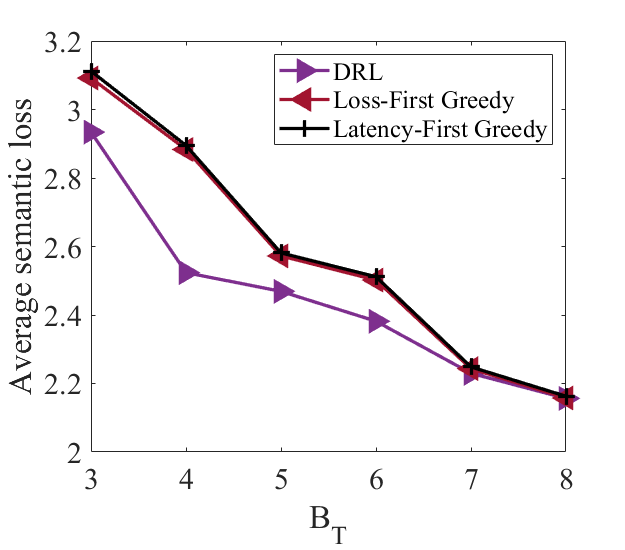}}
\subfigure[]{\includegraphics[width=0.24\textwidth]{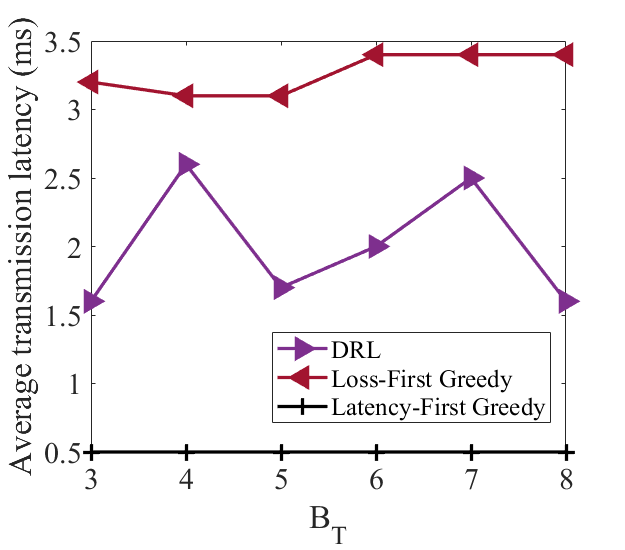}}
\subfigure[]{\includegraphics[width=0.24\textwidth]{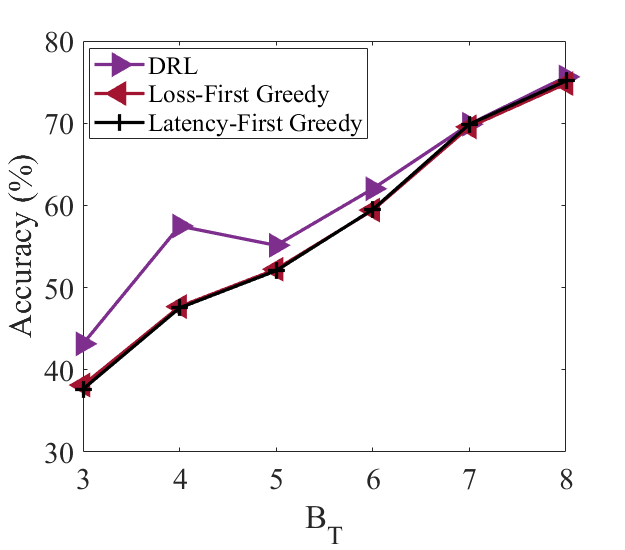}}
\caption{Effects of transmission latency constraint $\tau$ on (a) average semantic loss, (b) average transmission latency, (c) accuracy, and effects of transmitter SKB size $B_T$ on (d) average semantic loss, (e) average transmission latency, and (f) accuracy for DRL, Loss-First Greedy, and Latency-First Greedy algorithms.}
  \label{fig:vary_tau}
\end{figure}

Fig.~\ref{fig:vary_tau}(d)-(f) illustrates the average semantic loss, average transmission latency, and accuracy of the three algorithms under different transmitter SKB sizes $B_T$. The DRL consistently outperforms the two baselines, achieving significantly smaller average semantic loss and higher accuracy across all setups. As $B_T$ increases, the semantic loss for all three algorithms decreases (see Fig.~\ref{fig:vary_tau}(d)), and the classification accuracy increases (see Fig.~\ref{fig:vary_tau}(f)). This improvement is attributed to the larger transmitter SKB ensuring that better information and features can be transmitted on the transmitter side within the same latency constraints. However, the performance gap between the DRL and the two greedy algorithms narrows as $B_T$ increases. This is because with a large transmitter SKB, transmitting the lowest-level estimated class yields good results with minimal semantic loss and the shortest latency. As a result, the optimal actions for both the DRL and Loss-First Greedy algorithm gradually converge to those of the Latency-First Greedy algorithm, which consistently chooses the 4th-level transmission.

\section{Conclusion}
In summary, this study presents an innovative E2E communication framework designed for remote object recognition. Leveraging a semantic knowledge base and multi-level feature transmission and incorporating a novel online optimization model, our approach addresses the challenges posed by dynamic channel conditions and mobile device mobility. The proposed SAC-based DRL solver ensures minimal semantic loss and compliance with latency constraints. Simulation results showcase the superiority of this approach over conventional baselines, indicating significant potential for enhancing edge inference. Future work may further explore the system's performance under diverse setups and in more complex scenarios.

\bibliographystyle{IEEEtran}
\bibliography{bibtex}

\end{document}